\pdfoutput=1
\documentclass[aps,pre,longbibliography,showkeys,twocolumn,superscriptaddress]{revtex4-1}

\usepackage[english]{babel}
\usepackage{color}
\usepackage{listings}
\usepackage{float}
\usepackage{graphicx}
\usepackage{amsmath}
\usepackage{placeins}
\usepackage{hyperref}

\usepackage{braket}
\usepackage{bbm}
\usepackage{seqsplit}
\usepackage{booktabs,array}
\usepackage{tikz}
\usepackage{varwidth}
\usepackage{enumitem}
\usepackage{xparse}
\usepackage{etoolbox}
\usetikzlibrary{patterns,shadows.blur}

\graphicspath{{.},{./figures},{../figures}}

\usepackage{listings}

\definecolor{dblue}{rgb}{0,0,0.7}
\definecolor{lgray}{rgb}{.4,.4,.4}
\usepackage[framemethod=TikZ]{mdframed}

\begin{document}

\title{Advantages of a modular high-level quantum programming framework}

\author{Damian S. Steiger}
\email{dsteiger@phys.ethz.ch}
\affiliation{Theoretische Physik, ETH Zurich, 8093 Zurich, Switzerland}
\author{Thomas H\"aner}
\email{haenert@phys.ethz.ch}
\affiliation{Theoretische Physik, ETH Zurich, 8093 Zurich, Switzerland}
\author{Matthias Troyer}
\affiliation{Theoretische Physik, ETH Zurich, 8093 Zurich, Switzerland}

\begin{abstract}
	We review some of the features of the ProjectQ software framework and quantify their impact on the resulting circuits. The concise high-level language facilitates implementing even complex algorithms in a very time-efficient manner while, at the same time, providing the compiler with additional information for optimization through code annotation -- so-called \textit{meta-instructions}. We investigate the impact of these annotations for the example of Shor's algorithm in terms of logical gate counts. Furthermore, we analyze the effect of different intermediate gate sets for optimization and how the dimensions of the resulting circuit depend on a smart choice thereof. Finally, we demonstrate the benefits of a modular compilation framework by implementing mapping procedures for one- and two-dimensional nearest neighbor architectures which we then compare in terms of overhead for different problem sizes.
\end{abstract}

\keywords{Quantum Computing, Compilers, Quantum Programming Languages}

\maketitle

\section{Introduction}

Quantum computers will be able to solve certain problems faster than any classical supercomputers and thus enable finding solutions to problems that are intractable on any future classical computer. Quantum computers are not intended to replace classical technology. Rather, they should be viewed as special-purpose accelerators, similar to today's GPUs or FPGAs which are running in compute centers to speed up specific applications or subprocesses thereof. 

There are many reasons to believe that quantum computers will not replace classical computers. One is that most of the currently pursued technologies to build quantum bits require a vacuum chamber or temperatures on the order of milliKelvin, which makes them bulky and unsuitable for mobile technology. In addition, there are fundamental constraints for the programs running on quantum hardware due to the laws of quantum mechanics. In particular, all operations must be made reversible which incurs a large polynomial overhead in both space and time when translating a classical computation consisting of, e.g., NAND gates to reversible Toffoli gates. Furthermore, to successfully run a quantum program on a physical device, quantum error correction has to be employed in order to reduce the effects of noise on the computation. This causes quantum computers to run at a much lower clock speed than classical ones.

Hence, the focus of the quantum computing research community has been on finding applications for which a quantum algorithm has a large scaling advantage in time-to-solution, also known as a quantum speedup. A handful of such algorithms has been discovered such as the famous algorithm by Peter Shor for factoring integers~\cite{shor1994algorithms}. This algorithm scales super-polynomially better than the best known classical algorithm and has applications in breaking certain encryption schemes. So far only few examples of algorithms with quantum speedups are known and finding more is a very challenging task which is crucial to the development of the whole community.

\begin{figure}[t]
\resizebox{.8\linewidth}{!}{
\begin{tikzpicture}
\fill[pattern=north west lines,pattern color=red!20] (0,0) rectangle (2.99253,3.2);

\node (x1) at (5,0) {};
\node (y1) at (0,3.5) {};

\draw[->,thick] (0,0) -- (x1);
\draw[->,thick] (0,0) -- (y1);

\node (ylabel) at (-0.3,1.7) {\rotatebox{90}{\footnotesize Run time}};
\node (ylabel) at (2.4,-.25) {\rotatebox{0}{\footnotesize Problem size}};

\draw[thick,dashed,domain=0:4.7,smooth,variable=\x] plot ({\x},{0.145*\x*\x});
\draw[thick,domain=0:4.7,smooth,variable=\x] plot ({\x},{0.2*\x+0.7});

\node at (4.1,1.32) {\rotatebox{11.2}{\footnotesize quantum}};
\node at (3.85,2.45) {\rotatebox{48}{\footnotesize classical}};

\draw[draw,fill=red] (2.99253,0.2*2.99253+0.7) circle (2pt);
\draw[dashed,red,thick]  (2.99253,3.2) --  (2.99253,0);
\end{tikzpicture}
}
\caption{Illustration of what is typically encountered when comparing a quantum algorithm which exhibits a quantum speed-up to the best classical algorithm in terms of run time. The \textit{crossover point}, i.e., the problem size after which the quantum algorithm outperforms its classical counterpart, is shown as a red dashed line.}
\label{fig:crossoverpoint}
\end{figure}
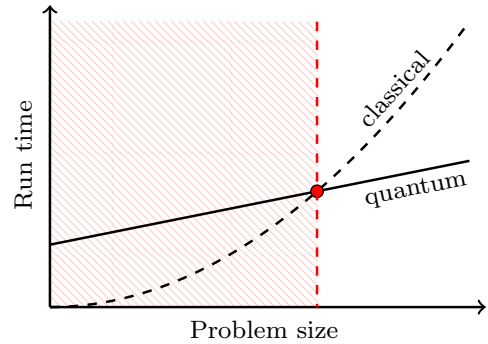

In order to determine if quantum algorithms with a scaling advantage can be useful for real applications, it is important to investigate at which problem size the \textit{crossover point} is reached after which the quantum algorithm has lower runtime, see Fig.~\ref{fig:crossoverpoint} for an illustration. If the crossover point is too far out, it might not be practical to use a quantum computer, e.g., if observing any speed advantage requires a runtime of at least the age of the universe~\cite{hhgtg}. Cost estimation of a quantum program can be achieved time-efficiently using a full stack software framework with a quantum programming language and sophisticated compilers. With these optimizing compilers, such a framework also allows to lower the crossover point even in the absence of large-scale quantum computers. This is crucial in order to leverage the economic potential of small-scale quantum computers as soon as possible.

In this paper, we are concerned with the software stack involved in running a quantum program. We provide a partial review of our software methodology~\cite{haener2016} which was then implemented resulting in the ProjectQ software framework for quantum computing~\cite{Steiger2018}, but with a new focus on some important aspects of the high-level programming language and new mappers. In particular, we show how different intermediate representations can decrease the quantum resources for the example of Shor's algorithm and quantify the resource improvements by using meta-instructions (code annotation) in the high-level language. We then introduce a new feature of ProjectQ, namely mapping to a linear chain of qubits with nearest neighbor gates or a two-dimensional square grid. Considering the overhead of mapping is important in determining the crossover points of quantum algorithms running on specific architectures. While our mappers scale optimally in terms of circuit depth, there is potential to reduce the constant factors by finding better heuristics. Providing these mappers and applications as open source software allows to incrementally improve their performance. Mappers are not just important in the long run, but  crucial in the current Noisy Intermediate-Scale Quantum (NISQ) technology era~\cite{nisq}, where quantum resources are very limited. We conclude with an outline of future research with and development of the ProjectQ framework.

\textbf{Related work}
Besides ProjectQ, there have been numerous other contributions in this field of quantum programming languages and compilers. A few of them are available as open source such as Quipper \cite{green2013quipper}, a quantum program compiler implemented in Haskell, the ScaffCC compiler based on the LLVM framework \cite{javadiabhari2014scaffcc}, IBM's QISKit \cite{qiskit}, and Rigetti's pyQuil \cite{pyquil}. Moreover, there are closed source quantum programming languages such as Microsoft's LIQ$Ui\ket{}$ \cite{wecker2014liquid} or Microsoft's Q\# \cite{svore2018q}, the latter of which currently allows executing the resulting circuits on a local simulator employing simulation kernels from ETH Zurich \cite{qsharplicense}.

The task of mapping a quantum circuit to a restricted interaction graph has been studied extensively. The aim of this paper is to provide model implementations of mappers in ProjectQ. Future work will be concerned with extending and improving their performance beyond the current state of the art. Mapping to a linear nearest neighbor architecture is discussed in detail for example in \cite{saeedi2011synthesis, hirata2011, shafaei2013optimization}. Our implementation is similar to Hirata \emph{et al.}~\cite{hirata2011} for a linear nearest neighbor architecture. Our algorithm finds qubit placements using a greedy search while theirs also employs more compute-intensive optimizations in order to reduce the total number of swaps. We improve upon their method by using a standard odd-even transposition sort \cite{habermann1972parallel} instead of bubble sort for the routing which can reduce the circuit depth by a constant factor using the same number of swaps. Our implementation of a mapper for the two-dimensional square grid follows the description in \cite{Brierley2017, Szegedy2017}. It has been known since 1986 that there are sorting network for square grids which have a worst-case overhead of $3\sqrt{n}$ in circuit depth for a grid with $n$ points, see \cite{Schnorr1986}.

\section{Compilation to quantum hardware}

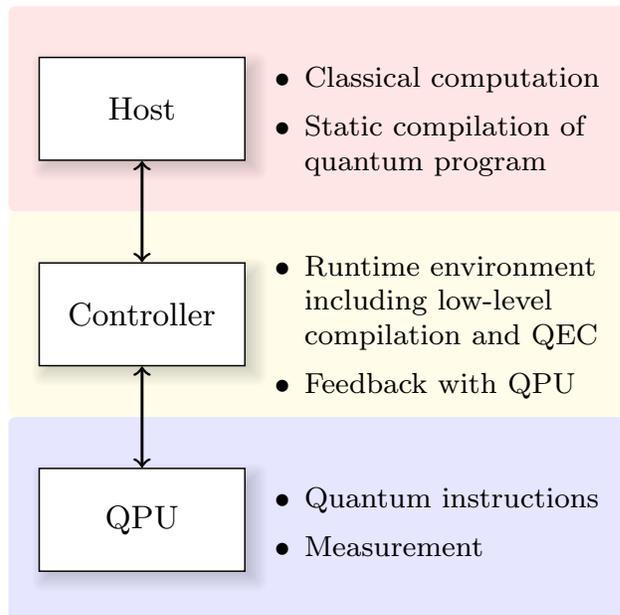
\begin{figure}[t]
	\resizebox{\linewidth}{!}{
		\begin{tikzpicture}
			\tikzstyle{basicshadow}=[blur shadow={shadow blur steps=20, shadow xshift=2pt, shadow yshift=-2pt, shadow scale=1.04, shadow opacity=20, shadow blur radius=0.8ex}]
			\fill[color=red!10, rounded corners=2pt] (-2,0.5) rectangle (4,-1.5);
			\fill[color=yellow!10, rounded corners=2pt] (-2,-1.5) rectangle (4,-3.5);
			\fill[color=blue!10, rounded corners=2pt] (-2,-3.5) rectangle (4,-5.5);
			
			\node[basicshadow,fill=white,draw,minimum width=2cm,minimum height=1cm] (pc) at (-.7,-.5) {Host};
			\node[basicshadow,fill=white,draw,minimum width=2cm,minimum height=1cm] (llpc) at (-.7,-2.5) {Controller};
			\node[basicshadow,fill=white,draw,minimum width=2cm,minimum height=1cm] (qpu) at (-.7,-4.5) {QPU};
			\draw[<->,thick] (pc)--(llpc);
			\draw[<->,thick] (llpc)--(qpu);
			\node[text width=4cm,font=\footnotesize] at (2,-.5) {
				\begin{itemize}
					\item Classical computation
					\item Static compilation of quantum program
				\end{itemize}};
			\node[text width=4cm,font=\footnotesize] at (2,-2.5) {
				\begin{itemize}
					\item Runtime environment including low-level compilation and QEC
					\item Feedback with QPU
				\end{itemize}};
			\node[text width=4cm,font=\footnotesize] at (2,-4.4) {
				\begin{itemize}
					\item Quantum instructions
					\item Measurement
				\end{itemize}};
		\end{tikzpicture}
	}
	\caption{Different levels of logic in a quantum hardware stack. For some architectures, the number of intermediate hardware levels may vary and lower parts of the stack may reside in a cryostat \cite{hornibrook2015cryogenic}.}
	\label{fig:hardware}
\end{figure}

The goal of a quantum software stack is to compile quantum programs to run them on actual quantum hardware. We give a short overview of a quantum software stack, outlining the challenges involved from a high-level perspective before going into the details in the next sections. For a recent review on quantum programming, see Ref.~\cite{chong2017programming} and more details on our methodology can be found in~\cite{haener2016}.

A high-level schematic of a large-scale quantum computer is shown in Fig.~\ref{fig:hardware}. A quantum computer functions as an accelerator for a classical host computer to solve specific subproblems. The software stack running on the host computer performs the static compilation of a quantum program. This process includes decomposing operations into a low-level logical gate set such as, e.g., the two-qubit CNOT gate and single-qubit rotations. After decomposing the quantum program into a low-level gate set, the compiler on the host computer has to map all operations to a restricted connectivity graph where, e.g., only nearest-neighbor qubits can perform a CNOT gate and hence qubits may have to be routed by swap operations.

Closer to hardware, we imagine a powerful classical controller which provides the runtime software environment. This includes error correction, rotation synthesis, and magic state distillation \cite{campbell2017roads}. Note that while accelerators such as GPUs are at least partially independent of the host computer, i.e., they have their own operation fetch mechanisms, a quantum computer requires that a classical chip dictates each operation to be executed.

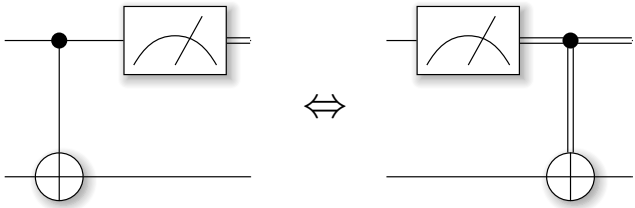
\begin{figure}[t]
	\resizebox{\linewidth}{!}{\begin{tikzpicture}[scale=0.8, transform shape]

\tikzstyle{basicshadow}=[blur shadow={shadow blur steps=8, shadow xshift=0.7pt, shadow yshift=-0.7pt, shadow scale=1.02}]\tikzstyle{basic}=[draw,fill=white]
\tikzstyle{operator}=[basic,basicshadow,minimum size=1.5em]
\tikzstyle{phase}=[fill=black,shape=circle,minimum size=0.1cm,inner sep=0pt,outer sep=0pt,draw=black]
\tikzstyle{none}=[inner sep=0pt,outer sep=-.5pt,minimum height=0.5cm+1pt]
\tikzstyle{measure}=[operator,inner sep=0pt,minimum height=0.5cm, minimum width=0.75cm]
\tikzstyle{xstyle}=[circle,basic,basicshadow,minimum height=0.35cm,minimum width=0.35cm,inner sep=-1pt,very thin]
\tikzset{
shadowed/.style={preaction={transform canvas={shift={(0.5pt,-0.5pt)}}, draw=gray, opacity=0.4}},
}
\tikzstyle{swapstyle}=[inner sep=-1pt, outer sep=-1pt, minimum width=0pt]
\tikzstyle{edgestyle}=[very thin]

\node[none] (line0_gate0) at (0.7,-0) {};
\node[none] (line1_gate0) at (0.7,-1) {};
\node[xstyle] (line1_gate1) at (1.1,-1) {};
\draw[edgestyle] (line1_gate1.north)--(line1_gate1.south);
\draw[edgestyle] (line1_gate1.west)--(line1_gate1.east);
\node[phase] (line0_gate1) at (1.1,-0) {};
\draw (line0_gate1) edge[edgestyle] (line1_gate1);
\draw (line1_gate0) edge[edgestyle] (line1_gate1);
\draw (line0_gate0) edge[edgestyle] (line0_gate1);
\node[measure,edgestyle] (line0_gate2) at (1.95,-0) {};
\draw[edgestyle] ([yshift=-0.18cm,xshift=0.075cm]line0_gate2.west) to [out=60,in=180] ([yshift=0.035cm]line0_gate2.center) to [out=0, in=120] ([yshift=-0.18cm,xshift=-0.075cm]line0_gate2.east);
\draw[edgestyle] ([yshift=-0.18cm]line0_gate2.center) to ([yshift=-0.075cm,xshift=-0.18cm]line0_gate2.north east);
\draw (line0_gate1) edge[edgestyle] (line0_gate2);
\node[none] (line0_gate3) at (2.5,-0) {};
\draw ([yshift=0cm]line0_gate3.center) edge [edgestyle] ([yshift=-0cm]line0_gate3.center);
\draw ([yshift=0.02cm]line0_gate2.east) edge[edgestyle] ([yshift=0.02cm]line0_gate3.west);
\draw ([yshift=-0.02cm]line0_gate2.east) edge[edgestyle] ([yshift=-0.02cm]line0_gate3.west);
\node[none] (line1_gate2) at (2.5,-1) {};
\draw ([yshift=0cm]line1_gate2.center) edge [edgestyle] ([yshift=-0cm]line1_gate2.center);
\draw (line1_gate1) edge[edgestyle] (line1_gate2);

\node at (3.05cm,-.5cm) {$\Leftrightarrow$};
\begin{scope}[yshift=2cm, xshift=2.8cm]

\node[none] (line2_gate0) at (0.7,-2) {};
\node[measure,edgestyle] (line2_gate1) at (1.3,-2) {};
\draw[edgestyle] ([yshift=-0.18cm,xshift=0.075cm]line2_gate1.west) to [out=60,in=180] ([yshift=0.035cm]line2_gate1.center) to [out=0, in=120] ([yshift=-0.18cm,xshift=-0.075cm]line2_gate1.east);
\draw[edgestyle] ([yshift=-0.18cm]line2_gate1.center) to ([yshift=-0.075cm,xshift=-0.18cm]line2_gate1.north east);
\draw (line2_gate0) edge[edgestyle] (line2_gate1);
\node[none] (line3_gate0) at (0.7,-3) {};
\node[xstyle] (line3_gate1) at (2.05,-3) {};
\draw[edgestyle] (line3_gate1.north)--(line3_gate1.south);
\draw[edgestyle] (line3_gate1.west)--(line3_gate1.east);
\node[phase] (line2_gate2) at (2.05,-2) {};
\draw ([xshift=0.02cm]line2_gate2.south) edge[edgestyle] ([xshift=0.02cm]line3_gate1.north);
\draw ([xshift=-0.02cm]line2_gate2.south) edge[edgestyle] ([xshift=-0.02cm]line3_gate1.north);
\draw (line3_gate0) edge[edgestyle] (line3_gate1);
\draw ([yshift=0.02cm]line2_gate1.east) edge[edgestyle] ([yshift=0.02cm]line2_gate2.west);
\draw ([yshift=-0.02cm]line2_gate1.east) edge[edgestyle] ([yshift=-0.02cm]line2_gate2.west);
\node[none] (line2_gate3) at (2.50,-2) {};
\draw ([yshift=0cm]line2_gate3.center) edge [edgestyle] ([yshift=-0cm]line2_gate3.center);
\draw ([yshift=0.02cm]line2_gate2.east) edge[edgestyle] ([yshift=0.02cm]line2_gate3.west);
\draw ([yshift=-0.02cm]line2_gate2.east) edge[edgestyle] ([yshift=-0.02cm]line2_gate3.west);
\node[none] (line3_gate2) at (2.50,-3) {};
\draw ([yshift=0cm]line3_gate2.center) edge [edgestyle] ([yshift=-0cm]line3_gate2.center);
\draw (line3_gate1) edge[edgestyle] (line3_gate2);
\end{scope}
\end{tikzpicture}}
	\caption{If fast feedback is available, quantum controls before measurement can be turned into classical controls after measurement. The opposite transformation can be applied if feedback is slow. The same idea carries over to other control instructions such as loops.}
	\label{fig:cnotmeasure}
\end{figure}

The software stack of today's quantum hardware is significantly simpler than the above because existing devices feature only a few tens of qubits and there is not yet a distinction between physical and logical qubits. As a consequence, the current experimental setups do not yet require a powerful classical controller and runtime software environment. However, it is possible to start exploring optimization opportunities in some technologies where, e.g., fast measurement feedback is possible. A simple example is shown in Fig.~\ref{fig:cnotmeasure}, where fast feedback can be used to measure a qubit earlier and hence reduce the effects of decoherence if it is possible to apply quantum operations depending on the measurement outcome. In this paper we will not focus on the classical controller part and the runtime software environment. However, when designing a software framework such as ProjectQ, it is important to be aware of these upcoming changes in order to design the framework accordingly. We discuss what types of interfaces are required from the host computer to the classical controller and how they can be added to ProjectQ and its high-level language.

Compilation and resource estimation of large quantum algorithms are limited due to performance bottlenecks in some compilers. This can become a problem already today, e.g., when trying to determine crossover points. On the other hand, compilation for current hardware is still sufficiently fast because noise limits quantum programs to circuit depths of below 100 gate operations. Furthermore, since current technologies still support arbitrary single-qubit rotations due to the absence of a quantum error correction protocol, these rotations do not need to be synthesized yet. As a consequence, the first quantum programs contain very few operations and the compiler running on the host computer will not exhibit any performance bottlenecks. As an example, while the decision problem whether we can map a quantum circuit to the connectivity graph of the underlying hardware in less than a specific circuit depth increase is $\mathsf{NP}$-complete~\cite{banerjee2016routing}, for general graphs, we can still find a close to minimal circuit-depth overhead solution using a brute-force approach for near-term hardware. It is important, however, to keep in mind that we should only apply compilation techniques which at least scale to quantum program sizes that we cannot classically simulate because only there, a quantum computer might show an advantage. All smaller programs are just proof of concept along the way toward larger quantum computers.

\section{The ProjectQ framework}

ProjectQ is a full stack, open source software framework which is implemented as an embedded domain-specific language (eDSL) in Python. For an introduction we refer the reader to our release paper~\cite{Steiger2018} and the code examples and tutorials which are available online~\cite{projectq}. ProjectQ defines a high-level language and compiles quantum programs to various backends, including quantum hardware such as the IBM Quantum Experience chips. 

To support research in quantum computing, we also bundle various software backends and analysis tools into our framework such as a resource counter, which provides performance information such as the number of gates and circuit depth of the compiled programs.
Moreover, for some of the proposed quantum algorithms, for example the variational quantum eigensolver~\cite{Peruzzo2014}, the success probability and/or the scaling with problem size are known asymptotically at best. Therefore, we also require scalable quantum simulators in order to run small quantum programs and extrapolate the performance in the absence of noise. Besides this, quantum simulators allow to find bugs in quantum code in a very pragmatic way. While one may want to aim at proving a program to be correct, the complexity of quantum programs is even higher than of classical distributed programs for which we currently fail to verify even small subroutines such as, e.g., certain locks. As a consequence, we do not expect to be able to theoretically prove the correctness of every quantum program. Rather, we envision a combination of theoretic validation and pragmatic testing to be the approach of choice. Detailed information about our quantum simulator can be found in Ref.~\cite{Steiger2018} and we discuss a highly scalable distributed quantum simulator in Ref.~\cite{haener2017}. Because quantum programs in ProjectQ are written in a high-level language, we can furthermore use emulation techniques to significantly speed up the quantum simulator for some algorithms~\cite{Haner2016Emulator}.

\subsection{High-level language}
We first consider the levels of abstractions in a quantum programming language. Currently there are two main levels of abstractions used in the quantum computing community.
On the one hand, quantum algorithm researchers work at the highest level of abstraction, where algorithms and subroutines are often specified in terms of their complexity in big $\mathcal O$ notation. This notation does not take into account constant factors which are important to determine crossover points with classical algorithms. On the other hand, researchers which are closer to experiments are implementing small quantum algorithms in the native gates of their hardware technologies. This is in line with recent open source programs introducing so-called quantum assembly languages \cite{cross2017qasm}.
Writing quantum programs on such a low level has the benefit of optimally using the available quantum hardware but also comes with the downsides which are encountered in classical computing: it makes writing useful programs a very time-consuming task and the resulting programs are not portable. A high-level programming language, on the other hand, has the advantages of shorter development time, less burden for the programmer to understand all details, and code portability. In our software framework ProjectQ, we provide both approaches, similar to what is done in classical high-performance computing today, where programs are written in high-level languages such as Python or C++ but certain performance bottlenecks are written as inline assembly code. Of course one then partially loses portability. However, we envision the quantum software stack as both application- and hardware-specific. Different hardware-specific functions together with a generic implementation, which works on all system, can be packaged into a library.
For the programmer, a high-level language combined with low-level instructions has the advantage that one does not need to learn a different language for writing application code or implementing a hand-optimized library function specific to a certain hardware technology. It is up to the good judgment of the programmer to choose the right level of abstraction for the task in question.

As an example, considern the following code written in the ProjectQ language:

\begin{lstlisting}[numbers=none,xleftmargin=21pt,framexleftmargin=17pt, label=code:syntax]
CNOT | (control_qubit, qubit)
MultiplyByConstantModN(a, N) | quint
\end{lstlisting}
It is inspired by the bra-ket notation used in physics, i.e., $ U \ket{\psi}$, where $U$ is a unitary operation applied to a wavefunction $\psi$. We use the \emph{or} operator  ($|$) of Python to achieve a similar syntax. It helps distinguishing our eDSL statements from normal Python code and separates the \emph{operation} with classical parameters on the left from the \emph{qubits} on the right side of the \emph{or} operator. In our language, we call an operation applied to specific qubits a \emph{command}.
The first command is a low-level controlled not operation acting on two qubits, while the second command is a high-level multiplication by a constant $a$ modulo $N$ applied to a register of quantum bits which are interpreted as a quantum integer.

\subsubsection*{Meta-instructions}

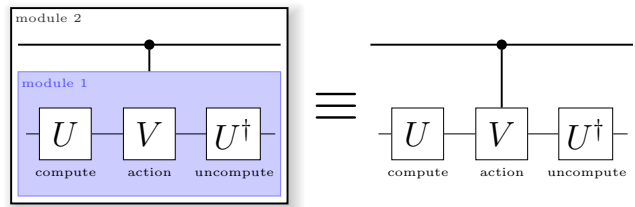
\begin{figure}[!t]
	\resizebox{\linewidth}{!}{
	\begin{tikzpicture}[x=1cm,y=1cm]
	\tikzstyle{basicshadow}=[blur shadow={shadow blur steps=20, shadow xshift=2pt, shadow yshift=-2pt, shadow scale=1.04, shadow opacity=20, shadow blur radius=0.8ex}]
	\draw[thick,basicshadow,fill=white] (0,.13) rectangle (4,-2.7);
	\node[text=black!80] at (0.54,-.01) {\tiny{module 2}};
	
	\begin{scope}[shift={(0.1,-0.4)},x=3.8cm,y=3.8cm]
		\draw[thick] (0,0)--(1,0);
		
		\node[draw,color=blue!50,fill=blue!20,minimum width=3.8cm,minimum height=1.8cm] (mod1) at (0.5, -0.34) {};
		\node[fill=black,circle,minimum size=0.15cm,inner sep=0pt,outer sep=0pt] (ctrl) at (0.5,0) {};
		\draw[thick] (ctrl) --(mod1);
		
		\begin{scope}[shift={(0.18,-0.14)}]
		\node[text=blue!70] at (-.04,-.004) {\tiny{module 1}};
		\node (O) at (-.18,-.2) {};
		\node[draw,fill=white,minimum height=.75cm,minimum width=.75cm] (U) at (0,-.2) {\Large $U$};
		\draw (O) --(U);
		\node[draw,fill=white,minimum height=.75cm,minimum width=.75cm] (V) at (.32,-.2) {\Large $V$};
		\draw (U) --(V);
		\node[draw,fill=white,minimum height=.75cm,minimum width=.75cm] (Udag) at (.64,-.2) {\Large $U^\dagger$};
		\draw (V) --(Udag);
		\node (e) at (0.83,-.2) {};
		\draw (Udag)--(e);
		\node at (.34,-.35) {\tiny{compute\hspace{.45cm}action\hspace{.33cm}uncompute}};
		\end{scope}
	\end{scope}
	
	\node at (4.7,-1.3) {\Huge $\equiv$};
		
	\begin{scope}[shift={(5.2,-0.4)},x=3.8cm,y=3.8cm]
		\draw[thick] (0,0)--(1,0);
		\node[fill=black,circle,minimum size=0.15cm,inner sep=0pt,outer sep=0pt] (ctrl) at (0.5,0) {};

		\begin{scope}[shift={(0.18,-0.14)}]
		\node (O) at (-.18,-.2) {};
		\node[draw,fill=white,minimum height=.75cm,minimum width=.75cm] (U) at (0,-.2) {\Large $U$};
		\draw (O) --(U);
		\node[draw,fill=white,minimum height=.75cm,minimum width=.75cm] (V) at (.32,-.2) {\Large $V$};
		\draw (U) --(V);
		\node[draw,fill=white,minimum height=.75cm,minimum width=.75cm] (Udag) at (.64,-.2) {\Large $U^\dagger$};
		\draw (V) --(Udag);
		\node (e) at (0.83,-.2) {};
		\draw (Udag)--(e);
		
		\node at (.34,-.35) {\tiny{compute\hspace{.45cm}action\hspace{.33cm}uncompute}};
		
		\draw[thick] (ctrl) --(V);
		\end{scope}
	\end{scope}
	
	\end{tikzpicture}
	}
	\caption{Submodule which contains a \emph{Compute/Action/Uncompute} pattern executed by a higher level module conditional on a qubit being in state 1. In general, if a submodule is run controlled on a qubit, the compiler has to control each operation in module 1 on the control qubit of module 2 being in state 1. However, in this scenario the compiler only needs to control the operation $V$ as the other two operations $U$ and $U^{\dagger}$ result in the identity if $V$ is not applied.}
	\label{fig:compute_action_uncompute}
\end{figure}

As a high-level language, ProjectQ contains many modules which are hierarchical combinations of lower-level subroutines. However, a simple concatenation of different circuit blocks would yield suboptimal constructs with a tremendous overhead as we will see in this and the next subsection. Fortunately, it can be avoided using code annotation in the high-level language together with an optimizing compiler as seen in the next subsection. In combination, these two features allow to drastically reduce the quantum resource requirements of a given implementation.

\begin{figure}[t]
	\resizebox{\linewidth}{!}{\input{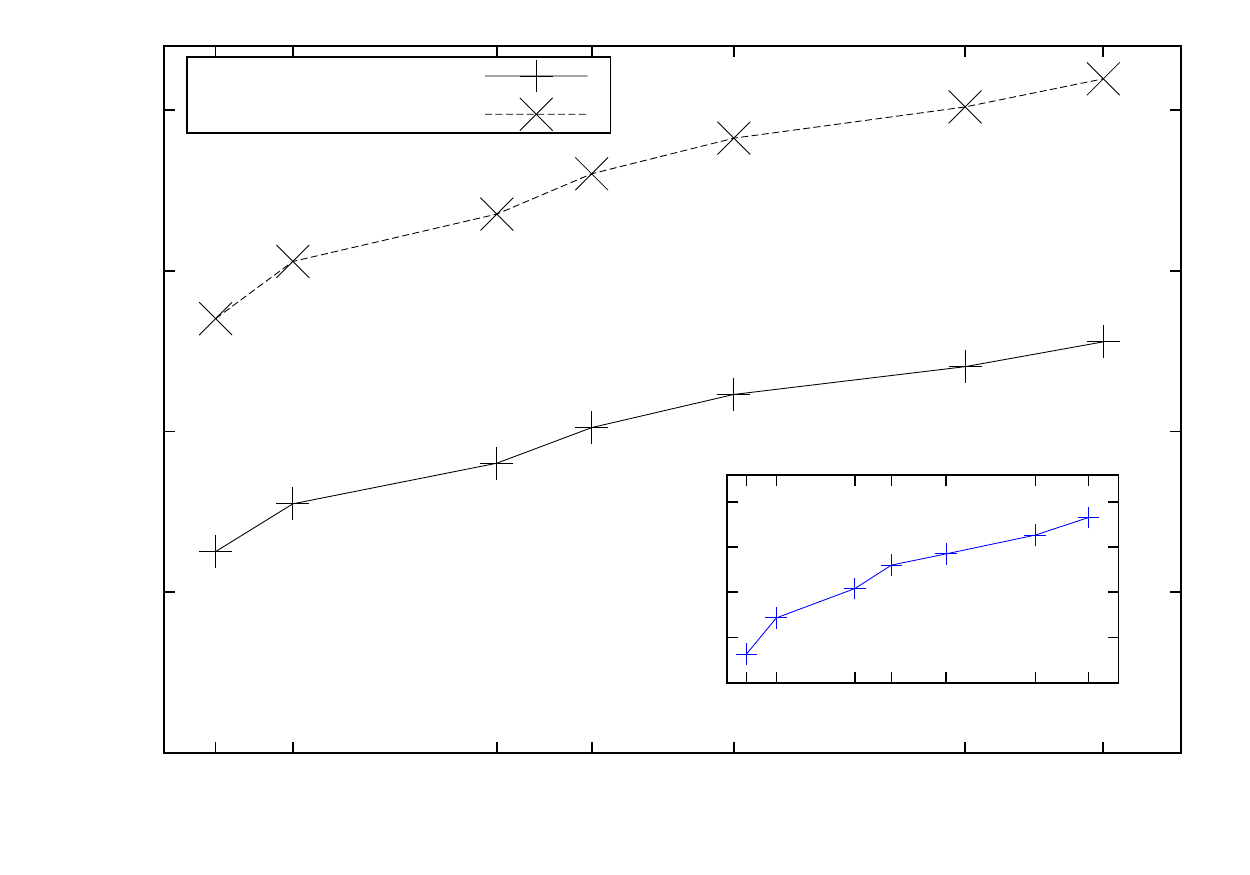}}
	\caption{Comparison of the number of CNOT gates which result from compilation with and without special handling of compute (C) / uncompute (UC) sections that allows for better optimization of subroutines which are executed controlled on other qubits. The data without C/UC for $N\geq 77$ was extrapolated using the CNOT gate counts of the first iteration of Shor's algorithm with a multiplier of $2\lceil\log_2 N\rceil$, which is the total number of iterations in the iterative quantum phase estimation as implemented in~\cite{Beauregard:2003:CSA:2011517.2011525}. Since each iteration features a controlled modular multiplication, the difference in gate counts between iterations is minor and results only from the fact that the multiplication is carried out by a different constant.}
	\label{fig:controlcnot}
\end{figure}

We now consider the combination of two very common design patterns in quantum programs: controlled execution of a subroutine conditioned on the state of a qubit being 1 and a pattern which we called \emph{compute/action/uncompute} in \cite{haener2016}. The compute/action/uncompute pattern is just a sequence of three unitary operators $ U^{\dagger} V U$, where the first unitary operator is the inverse of third operator. This is very common when implementing classical functions reversibly on a quantum computer (using Bennett's trick \cite{Bennett1973}, where the first and third stage correspond to $U$ and $U^\dagger$, respectively) but also in quantum simulation. For examples see the ProjectQ quantum math library or the implementation of the \texttt{TimeEvolution} operator in ProjectQ. When combining these two patterns, there is a very simple optimization which can be done as shown in Fig.~\ref{fig:compute_action_uncompute}. For a compiler it would be extremely difficult to find these patterns as both $U$ and $V$ could feature several hundreds of operations. 
Therefore, we enable the programmer to annotate such design patterns for the compiler. In this particular case, we require that all gates are annotated with the information to which section of this design pattern (compute, action, or uncompute)  they belong. In ProjectQ, language constructs to annotate code with additional information are called meta-instructions. \newpage The syntax for this meta-instruction in ProjectQ is:

\begin{lstlisting}[numbers=none]
with Compute(eng):
    U | qureg
V | qureg
Uncompute(eng)
\end{lstlisting}

Our eDSL uses Python's context handler (\texttt{with ...}) to allow the programmer to specify $U$ as an indented block of instructions. Additionally, the context handler automatically creates the inverse $U^\dagger$ which is applied when calling the \texttt{Uncompute} function and hence this makes the code more compact and less error-prone. If this subroutine now gets executed conditional on some qubit being in state 1, the compiler can easily apply the optimization in Fig.~\ref{fig:compute_action_uncompute}. We demonstrate this using an implementation of Shor's algorithm, see Box~\ref{box:shor}. In order to maximize the potential for reuse of our implementation, we first implemented the required mathematical functions as gates in ProjectQ, thereby building a small math library for quantum computing. We calculated the resource overheads for small sizes of Shor's algorithm once with our compute/action/uncompute meta instruction enabled and once without. The results in Fig.~\ref{fig:controlcnot}  show a reduction of more than 40x in the number of CNOT gates. 

\begin{figure}
	\begin{mdframed}[
		frametitle={Shor's algorithm},
		outerlinewidth=0.6pt,
		innertopmargin=6pt,
		innerbottommargin=6pt,
		roundcorner=4pt]
		Shor's algorithm allows to find the prime factors of a $n$-bit number $N$.
		We use an implementation that is based on Beauregard's 2$n$+3 qubit circuit for Shor's algorithm~\cite{Beauregard:2003:CSA:2011517.2011525} but instead of implementing the entire circuit following the paper, we build a math library which we then use to implement the modular exponentiation routine which achieves
		\[
		\Ket x\Ket 0^{\otimes n}\mapsto \Ket x \Ket{ a^x\operatorname{mod} N}\;,
		\]
		where $a\in [2,...,N-1]$ is a randomly chosen integer, $N$ is the number to factor, and $\ket x$ is the input register consisting of $2n$ qubits in the uniform superposition state $\Ket x\propto\sum_i\Ket i$ with $n=\lceil\log_2 N\rceil$.
		After executing this modular exponentiation subroutine, an inverse quantum Fourier transform is applied, followed by a measurement of all input qubits. The output of this measurement can then be used to determine the period of $a^x\operatorname{mod} N$ and, in turn, the factors of $N$~\cite{shor1994algorithms}. The code can be found in our example algorithms \cite{projectq} or in the appendix of our ProjectQ paper \cite{Steiger2018}.
		
		There are several optimization opportunities and space/time tradeoffs which have been investigated in several works~\cite{Takahashi:2006:QCS:2011665.2011669,Haner:2017:FUQ:3179553.3179560,kutin2006shor,vanmetershor08,gidney2017factoring}. In the following, we shortly outline the most crucial optimizations performed by Beauregard. First, after decomposing the modular exponentiation of $a$ into modular multiplications by constants $a^{2^i}$, which can be implemented using modular additions, one can use Draper's addition in Fourier space \cite{draper2000addition} which requires no ancilla qubits to add a classical constant to a quantum register, thereby allowing to save $\mathcal O(n)$ qubits. More recent work also eliminates the substantial overhead from the quantum Fourier transform required for this type of adder by constructing a purely Toffoli-based network of depth $\mathcal O(n)$~\cite{Haner:2017:FUQ:3179553.3179560}. Second, using the circuit identity in Fig.~\ref{fig:cnotmeasure} from left to right, the final measurement gates can be pulled through parts of the inverse quantum Fourier transform, allowing to serialize the circuit for modular exponentiation such that only 1 of the $2n$ qubits of $\ket x$ need to be alive at the same time~\cite{PhysRevLett.76.3228}. One more qubit can be removed by improving the construction of the modular addition circuit\cite{Takahashi:2006:QCS:2011665.2011669} and recent work~\cite{gidney2017factoring} reduces the circuit width by another qubit to a total of $2n+1$.
	\end{mdframed}
	\renewcommand{\figurename}{Box}
	\caption{Shor's algorithm as used in this paper and discussion of algorithmic improvements which have been published.}
	\label{box:shor}
\end{figure}

\subsubsection*{Classical instructions for quantum computing}
We now consider how ProjectQ is currently dealing with classical functions and how it can be extended in the future. A quantum programming language does not just contain quantum gates but also classical operations.

There are two different kinds of classical instructions. First, there are classical functions which act on a superposition of inputs and thus must be executed on quantum hardware, an example being modular multiplication in Shor's algorithm. Second, there are classical control instructions which need to be executed by the classical controller, see Fig.~\ref{fig:hardware}. 

Many quantum algorithms require the execution of classical functions on a superposition of inputs and therefore, these classical functions have to be translated into reversible operations such as the Toffoli gate. ProjectQ already contains a small quantum math library which specifies such classical functions in our eDSL. For example,
\begin{minipage}{\linewidth}
	\begin{lstlisting}[numbers=none,xleftmargin=21pt,framexleftmargin=17pt, label=code:syntax]
	MultiplyByConstantModN(a, N) | quint
	\end{lstlisting}
\end{minipage}
Another approach was taken by RevKit which has recently been integrated into ProjectQ~\cite{soeken2018revkit}. Instead of extending our eDSL with all the classical math operations, it implements oracle functions in our eDSL which take a Python function as a parameter:
\begin{minipage}{\linewidth}
\begin{lstlisting}[numbers=none]
def classical_function(a,b,c,d):
    return (a and b) ^ (c and d)
PhaseOracle(classical_function) | qubits
\end{lstlisting}
\end{minipage}
When executed, our compiler traverses the AST of Python to get the definition of the classical function and then uses RevKit to synthesis a reversible version with Toffoli gates. For more information see the paper by Soeken \emph{et al.} \cite{soeken2018revkit}. 

Let us now turn to the second example where classical functions need to be executed by a classical controller which is located close to hardware. A standard example are classical control instructions such as loops or repeat-until-success constructs \cite{bocharov2015}. We have implemented such classical control flow instruction as meta-instructions which means they use the code annotation feature in our eDSL. For example, loops can be specified as:
\begin{minipage}{\linewidth}
\begin{lstlisting}[numbers=none]
with Loop(eng, 10):
   U | qubits
\end{lstlisting}
\end{minipage}
This will send the command to apply \texttt{U} to the \texttt{qubits} annotated with the classical control instruction to repeat it 10 times to the backend or if the backend does not support loops, the compiler will unroll it automatically. Similarly a repeat-until-success of a specific measurement outcome can be added to our eDSL. Because our programming language is embedded into Python, one can also use Python to perform loops or post-process measurement outcomes which then determine the next quantum operations. While this is currently possible when using a simulator as a backend, it is not possible when running the program on a real quantum device as these classical operations need to be executed on a controller because they require a lower latency to the quantum hardware. Our eDSL can be extended to handle such a scenario by introducing more elaborate classical meta-instructions:
\begin{minipage}{\linewidth}
\begin{lstlisting}[numbers=none]
with ClassicalControl(eng, function, qubits):
   # Execute this quantum code
\end{lstlisting}
\end{minipage}
where the classical function can either be specified in Python and then translated to, e.g., C or directly added as a string containing C code. Alternatively, we can add a new syntax keyword for kernel functions and use a custom pre-processor to extract these functions before the Python code is interpreted.

\subsection{Compiler design}

\begin{figure*}[!t]
	\includegraphics[width=\linewidth]{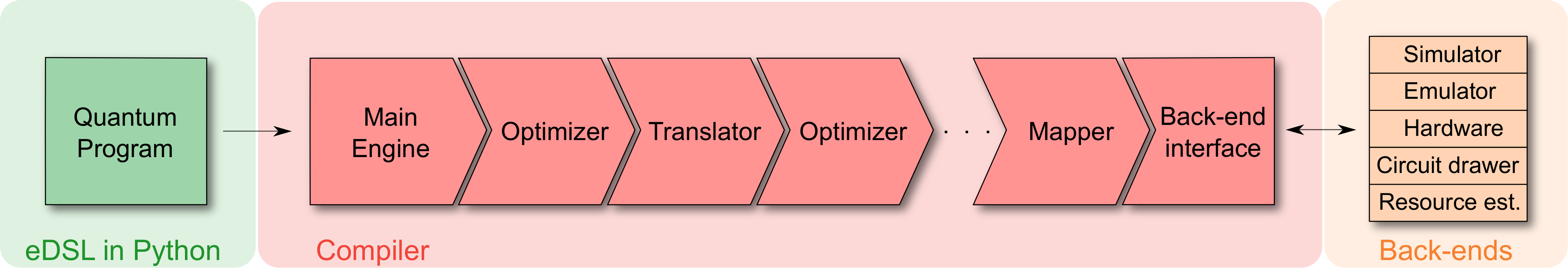}
	\caption{(reprinted from \cite{Steiger2018}) ProjectQ's full stack software framework. Users write their quantum programs in a high-level domain-specific language embedded in Python. The quantum program is then sent to the \texttt{MainEngine}, which is the front end of the modular compiler. The compiler consists of individual compiler engines which transform the code to the low-level instruction sets supported by the various back-ends, such as interfaces to quantum hardware, a high-performance quantum simulator and emulator, as well as a circuit drawer and a resource counter.}
	\label{fig:compilercomponents}
\end{figure*}

In order to keep as much code portability as possible using our mixed approach of high- and low-level instructions, ProjectQ is implemented as a modular framework which allows to adapt the compiler and intermediate representations to an application in order to better optimize the quantum program, see Fig.~\ref{fig:compilercomponents}. Individual compiler engines can be combined to make an application- and hardware-specific compiler in order to best use the limited quantum resources. In this section we will show how different intermediate representations can decrease the quantum resources for the example of Shor's algorithm. In addition, this modularity of the compiler allows for organic growth of the framework and its capabilities. While this happens, standardized interfaces which support several quantum architectures emerge.

\begin{figure}[t]
	\resizebox{\linewidth}{!}{\input{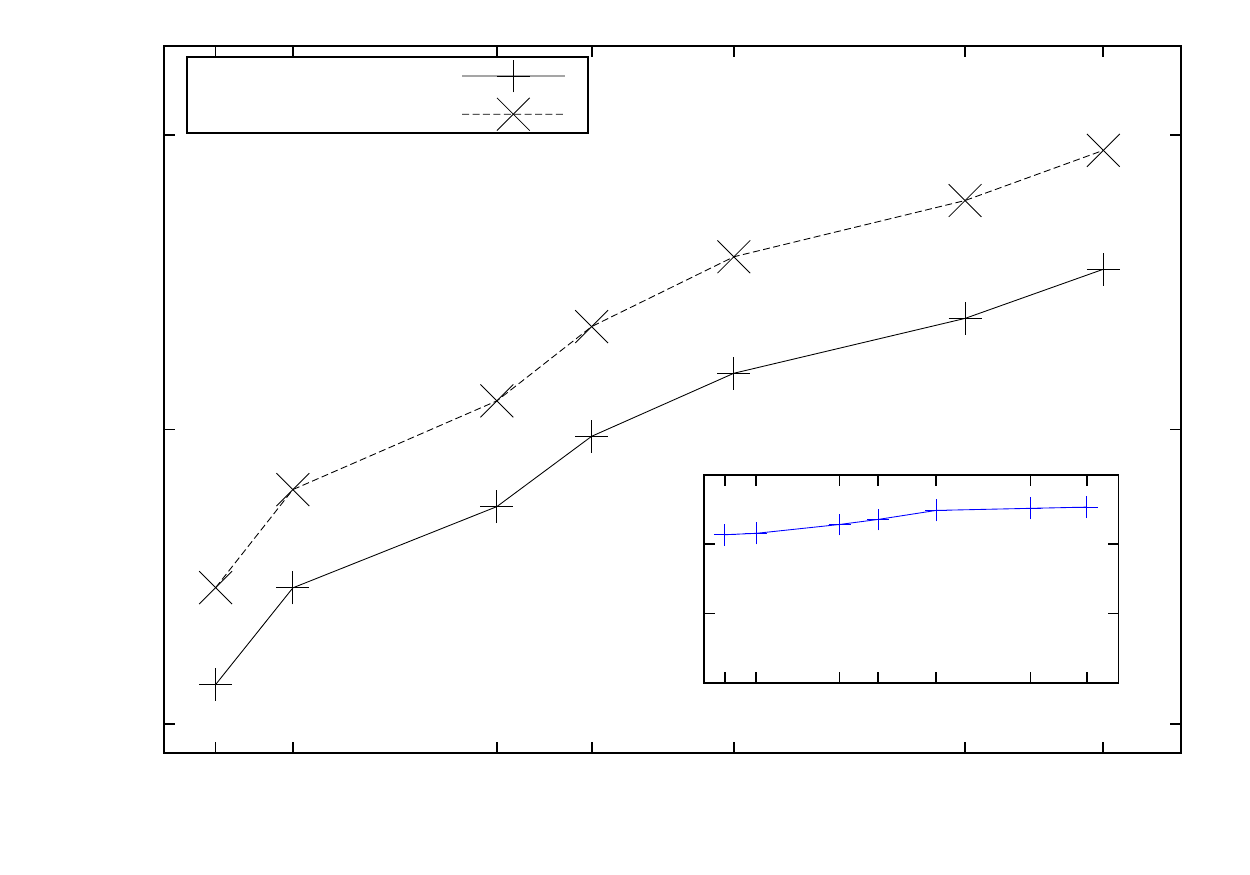}}
	\caption{Comparison between the number of Rz gates which result from compilation with and without an intermediate gate set (IGS), which allows for better optimization of the circuit. See the text for the definition of the gate set and optimization procedure.}
	\label{fig:intermrz}
\end{figure}

\begin{figure}[t]
	\resizebox{\linewidth}{!}{\input{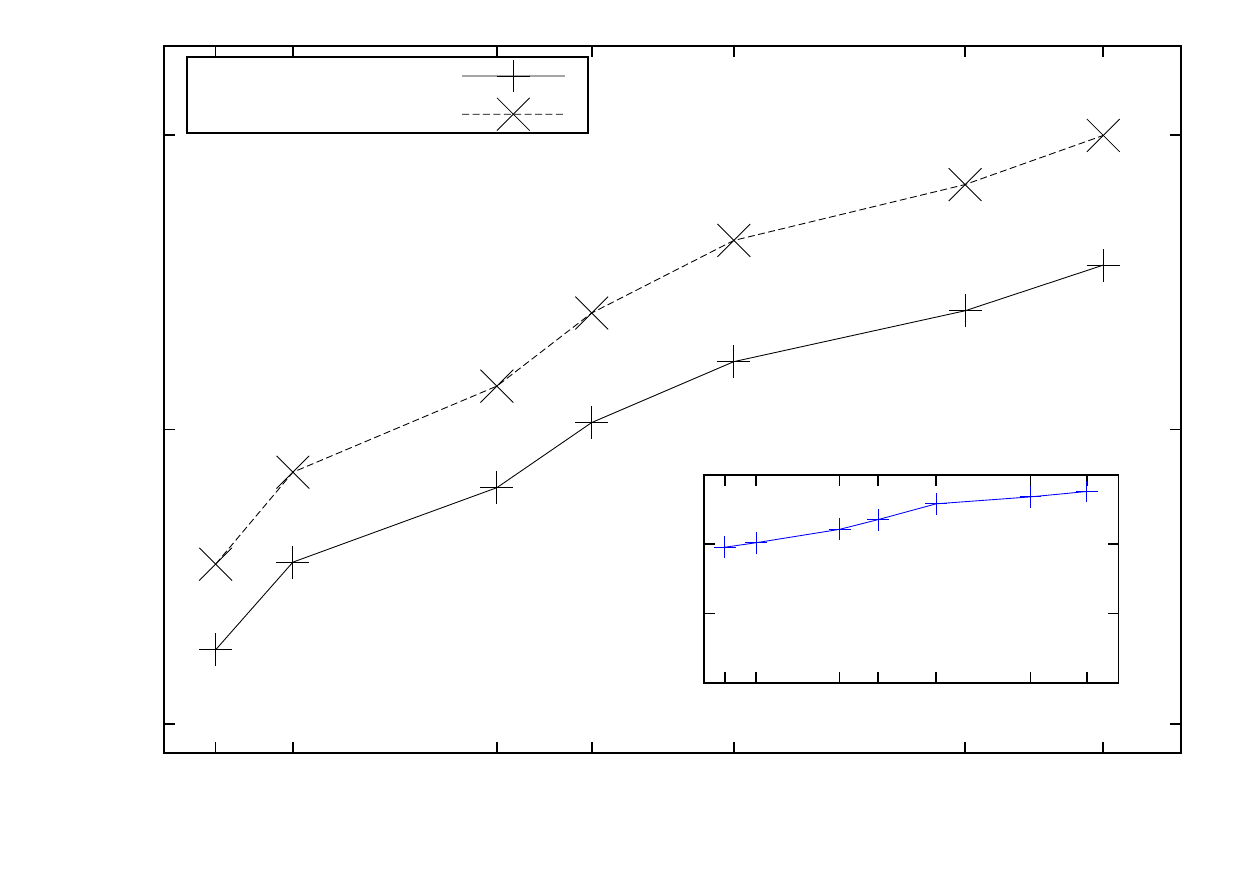}}
	\caption{Comparison between the number of CNOT gates which result from compilation with and without an intermediate gate set (IGS), which allows for better optimization of the circuit. See the text for the definition of the gate set and optimization procedure.}
	\label{fig:intermcnot}
\end{figure}

ProjectQ's compiler engines receive the quantum program in linear order and can transform the code before sending it on to the next engine. Because quantum programs become very large when translating them to low-level gates, it quickly becomes impossible for the host computer to store the entire program in memory. As a remedy, the compiler engines in ProjectQ only work on small parts of the code before sending it to the next engine and never require storage of the entire circuit.
Despite this locality, global optimizations are enabled through code annotations or automatic local optimizations at higher levels of abstraction. These optimizations can be made  more efficient by a smart choice of intermediate gate sets. Fig.~\ref{fig:intermrz} and Fig.~\ref{fig:intermcnot} show how this choice affects the resource requirements.
Shor's algorithm is compiled into a target gate set consisting of the two qubit CNOT gate and single qubit gates. To better differentiate the cost of single qubit gates for a fault tolerant quantum computer, we choose three categories: Clifford gates, T-gates, and Rz-gates. Using for example the standard surface code error correction scheme T-gates are more expensive due to magic state distillation and Rz-gates are even more expensive as they will require gate synthesis involving several T-gates \cite{campbell2017roads}. Our compiler did not perform any rotation synthesis so the T-gates originate only from the decomposition of Toffoli gates. We compiled Shor's algorithm using no intermediate gate set and hence the compiler first decomposes each operation into the target gate set before optimizations take place. In the second setting the compiler first decomposes the algorithm into an intermediate gate set (IGS) consisting of n-qubit QFT gates and arbitrary one- and two-qubit gates. We then perform an optimization step before decomposing into the target gate set at which point we perform another optimization round. With the IGS, the inverse QFT and QFT gate of two successive Draper addition circuits~\cite{draper2000addition} can be canceled easily. This is the main reason for the advantage of using an IGS in this example.

\subsection{Mapping quantum programs to hardware with limited connectivity}

A usual abstraction of high-level quantum programming languages is that operations can be performed on any set of qubits. This abstraction is useful as the programmer can focus on logical operations without having to worry about the underlying hardware constraints. Needless to say, it is never a bad idea if a programmer has some of the hardware constraints in mind when writing code. This is true for quantum programming as it is in the classical case, where having a good idea of the memory hierarchy can improve performance by many orders of magnitude.

In this section, we consider mapping a quantum program to a machine model which only allows single- and two-qubit gates, and where the latter can be executed only if the pair of qubits are neighbors on a given connectivity graph. Disjoint pairs of neighboring qubits can execute operations in parallel and we assume that the connectivity graph has low degree.
 In quantum computers, the connectivity graph is determined by the hardware and -- on future error corrected machines -- by the chosen error correction code. 
 For example, it is possible to build superconducting qubits on a linear nearest neighbor chain because this allows to use a two dimensional chip design with control lines coming from the sides. A technologically more advanced design is to have the control lines coming from the third dimension and hence allowing qubits to be connected on a nearest neighbor square grid \cite{2018interconnects}.
While for small experiments with tens of gates and only a few qubits the mapping process could be optimized manually, this will no longer be the case for hardware with more than 50 qubits especially as early devices might have irregular graphs due to faulty qubits. From the beginning, ProjectQ has been able to map small algorithms to the IBM Quantum Experience chip with 5 qubits. We have extended the interface to facilitate the implementation and optimization of mappers. This will allow to perform benchmarks for such mappers in the future using the same algorithms, e.g., Shor's algorithm which is already implemented in ProjectQ or quantum chemistry algorithms currently available in FermiLib \cite{projectq} or its fork OpenFermion \cite{openfermion}.

There are three different and competing performance metrics when optimizing mappers. First, a mapper might try to minimize the number of swap operations required to move qubits. This is a useful metric because swap operations can be implemented using, e.g., three CNOT gates which are noisy on today's hardware. Second, a mapper can try to reduce the increase in circuit depth by applying as many swap operations in parallel as possible because a lower circuit depth means faster run time. In principle, this also means that qubits require less coherence time but, on the other hand, the increase in swap operations might cancel out this effect. Third, one can increase the number of qubits in order to, e.g., keep the circuit depth unchanged up to a constant factor \cite{rosenbaum2012optimal}. 

We have implemented a mapper for a linear nearest neighbor chain topology and a two-dimensional square grid. For both mappers, we focus on reducing the circuit depth without increasing the number of qubits. Our mappers perform the mapping in three distinct phases:
\begin{enumerate}
\item Find a qubit placement on the hardware graph which puts interacting qubits next to each other
\item Route the qubits from the old positions to the new positions using swap operations
\item Apply all the operations which act on single or neighboring pairs of qubits
\end{enumerate}
This procedure is repeated until all commands have been executed. 

\begin{figure*}[t]
\resizebox{\linewidth}{!}{
	\begin{tikzpicture}[x=1cm,y=1cm]
	\edef\r{"red!20"}
	\edef\o{"orange!30"}
	\edef\y{"yellow!50"}
	\def\colors{{\y,\o,\y,\o,\y,\r,\o,\r,\r, \r,\o,\y,\r,\o,\y,\r,\o,\y, 			
				 \o,\r,\y,\o,\y,\r,\y,\o,\r, \r,\o,\y,\r,\o,\y,\r,\o,\y}}
	\def\numbers{{0,3,2,4,1,5,6,7,8, 7,6,2,8,3,1,5,4,0,
				  6,7,2,4,1,8,0,3,5, 7,6,2,8,4,1,5,3,0}}
	\def\plotnames{{"a)","d)","b)","c)",}}
	\def\yyshift{3.5cm}
	\def\xxshift{4cm}
	\foreach \xx in {0,1}
	{
	\foreach \yy in {0,1}
	{
		\begin{scope}[xshift=\xx*\xxshift,yshift=-\yy*\yyshift]
		\pgfmathparse{\plotnames[(\xx+\yy*2)]}
		\edef\nm{\pgfmathresult}
		\node at (-.75,.3) {\nm};
		\foreach \i in {0,1,2}
		{
		\pgfmathparse{\colors[(\xx*9+\yy*18+\i*3)]}
		\edef\clr{\pgfmathresult}
		\pgfmathparse{\numbers[(\xx*9+\yy*18+\i*3)]}
		\edef\nr{\pgfmathresult}
		\node[circle,draw,fill=\clr] (p1\i) at (0,-\i) {\nr};
		\pgfmathparse{\colors[(\xx*9+\yy*18+\i*3+1)]}
		\edef\clr{\pgfmathresult}
		\pgfmathparse{\numbers[(\xx*9+\yy*18+\i*3+1)]}
		\edef\nr{\pgfmathresult}
		\node[circle,draw,fill=\clr] (p2\i) at (1,-\i) {\nr};
		\pgfmathparse{\colors[(\xx*9+\yy*18+\i*3+2)]}
		\edef\clr{\pgfmathresult}
		\pgfmathparse{\numbers[(\xx*9+\yy*18+\i*3+2)]}
		\edef\nr{\pgfmathresult}
		\node[circle,draw,fill=\clr] (p3\i) at (2,-\i) {\nr};
		\draw (p1\i)--(p2\i);
		\draw (p2\i)--(p3\i);
		}
		\foreach \i in {0,1}
		{
		\pgfmathtruncatemacro{\iplusone}{\i+1}
		\draw (p1\i)--(p1\iplusone);
		\draw (p2\i)--(p2\iplusone);
		\draw (p3\i)--(p3\iplusone);
		}
		\end{scope}
	}
	}
	\draw[draw,green!70!black,thick] (-0.4cm,+0.5cm) rectangle (0.4cm,-2.5cm);
	\draw[draw,green!70!black,dashed,thick] (-0.4cm,-\yyshift+0.5cm) rectangle (0.4cm,-\yyshift-2.5cm);
	
	\draw[draw,green!90!black!30] (-0.4cm+1cm,+0.5cm) rectangle (0.4cm+1cm,-2.5cm);
	\draw[draw,green!90!black!30,dashed] (-0.4cm+1cm,-\yyshift+0.5cm) rectangle (0.4cm+1cm,-\yyshift-2.5cm);
		
	\draw[draw,green!90!black!30] (-0.4cm+2cm,+0.5cm) rectangle (0.4cm+2cm,-2.5cm);
	\draw[draw,green!90!black!30,dashed] (-0.4cm+2cm,-\yyshift+0.5cm) rectangle (0.4cm+2cm,-\yyshift-2.5cm);

	\draw[draw,red!10] (-0.5cm,-\yyshift+0.4cm) rectangle (+2.5cm,-\yyshift+0.4cm-0.8cm);
	\draw[draw,red!10] (-0.5cm,-\yyshift+0.4cm-1cm) rectangle (+2.5cm,-\yyshift+0.4cm-1.8cm);
	\draw[draw,red!30,thick] (-0.5cm,-\yyshift+0.4cm-2cm) rectangle (+2.5cm,-\yyshift+0.4cm-2.8cm);

	\draw[draw,red!10,dashed] (\xxshift-0.5cm,-\yyshift+0.4cm) rectangle (\xxshift+2.5cm,-\yyshift+0.4cm-0.8cm);
	\draw[draw,red!10,dashed] (\xxshift-0.5cm,-\yyshift+0.4cm-1cm) rectangle (\xxshift+2.5cm,-\yyshift+0.4cm-1.8cm);
	\draw[draw,red!30,thick,dashed] (\xxshift-0.5cm,-\yyshift+0.4cm-2cm) rectangle (\xxshift+2.5cm,-\yyshift+0.4cm-2.8cm);

	\draw[draw,blue!40,dashed] (-0.4cm+\xxshift,+0.5cm) rectangle (0.4cm+\xxshift,-2.5cm);
	\draw[draw,blue!40] (-0.4cm+\xxshift,-\yyshift+0.5cm) rectangle (0.4cm+\xxshift,-\yyshift-2.5cm);
	
	\draw[draw,blue,dashed,thick] (-0.4cm+\xxshift+1cm,+0.5cm) rectangle (0.4cm+\xxshift+1cm,-2.5cm);
	\draw[draw,blue,thick] (-0.4cm+\xxshift+1cm,-\yyshift+0.5cm) rectangle (0.4cm+\xxshift+1cm,-\yyshift-2.5cm);
	
	\draw[draw,blue!40,dashed] (-0.4cm+\xxshift+2cm,+0.5cm) rectangle (0.4cm+\xxshift+2cm,-2.5cm);
	\draw[draw,blue!40] (-0.4cm+\xxshift+2cm,-\yyshift+0.5cm) rectangle (0.4cm+\xxshift+2cm,-\yyshift-2.5cm);
	
	\draw[color=green!70!black,->,thick] (0,-2.6)--(0,-2.9);
	\draw[color=green!90!black!30,->] (1,-2.6)--(1,-2.9);
	\draw[color=green!90!black!30,->] (2,-2.6)--(2,-2.9);
	
	\draw[color=red!20,->] (\xxshift-1.15cm,-\yyshift-0.0cm)--(\xxshift-.85cm,-\yyshift-0.0cm);
	\draw[color=red!20,->] (\xxshift-1.15cm,-\yyshift-1.0cm)--(\xxshift-.85cm,-\yyshift-1.0cm);
	\draw[color=red!40,->,thick] (\xxshift-1.15cm,-\yyshift-2.0cm)--(\xxshift-.85cm,-\yyshift-2.0cm);
	
	\draw[color=blue!40,->] (\xxshift+0.02cm,-2.9)--(\xxshift+0.02cm,-2.6);
	\draw[color=blue,thick,->] (\xxshift+1.02cm,-2.9)--(\xxshift+1.02cm,-2.6);
	\draw[color=blue!40,->] (\xxshift+2.02cm,-2.9)--(\xxshift+2.02cm,-2.6);

	\begin{scope}[xshift=8cm,y=1.45cm,yshift=-1.45*0.8cm+.3cm,x=1cm]
	\node at (-1,.8) {e)};
	\pgfmathparse{\r}
	\edef\red{\pgfmathresult}
	\pgfmathparse{\o}
	\edef\orange{\pgfmathresult}
	\pgfmathparse{\y}
	\edef\yellow{\pgfmathresult}
	\node at (0,.5) {{initial columns}};
	\node at (4,.5) {{final columns}};
	
	\newcommand{\txt}[1]{}
	
	\node[fill=green!90!black!70,circle,thick] (01) at (0,0) {\textcolor{black}{$u_0$}};
	\node[fill=green!10,circle] (11) at (0,-1) {\textcolor{black}{$u_1$}};
	\node[fill=green!10,circle] (21) at (0,-2) {\textcolor{black}{$u_2$}};

	\node[fill=\red,circle,thick] (02) at (4,0) {\textcolor{black}{$v_0$}};
	\node[fill=\orange,circle,thick] (12) at (4,-1) {\textcolor{black}{$v_1$}};
	\node[fill=\yellow,circle,thick] (22) at (4,-2) {\textcolor{black}{$v_2$}};
	
	\draw[bend left=20,very thick] (01) to node [auto,near start,yshift=-0.1cm] {\txt{\textbf{\footnotesize 0}}} (12);
	\draw[bend left=-10,thin,dash pattern=on 4pt off 1pt] (01) to node [auto,near start,yshift=-0.1cm] {\txt{\footnotesize 1}} (12);
	\draw[bend left=-10,dotted,thin] (01) to node [auto,near start,swap,yshift=0.3cm,xshift=-0.2cm] {\txt{\footnotesize 2}} (22);
	
	\draw[bend right=8,very thick] (11) to node [auto,at end,yshift=-0.35cm,xshift=-0.4cm] {\txt{\textbf{\footnotesize 0}}} (02);
	\draw[bend left=-10,dotted,thin] (11) to node [auto,near end,yshift=-0.1cm,xshift=-.2cm] {\txt{\footnotesize 2}} (12);
	\draw[bend left=-10,thin,dash pattern=on 4pt off 1pt] (11) to node [auto,swap,near end,yshift=0.1cm] {\txt{\footnotesize 1}} (22);
	
	\draw[bend left=2,dotted,thin] (21) to node [auto,near start,yshift=-0.52cm,xshift=-0.4cm] {\txt{\footnotesize 2}} (02);
	\draw[bend left=-25,thin,dash pattern=on 4pt off 1pt] (21) to node [auto,swap,near start,yshift=-0.0cm,xshift=-0.3cm] {\txt{\footnotesize 1}} (02);
	\draw[bend left=-10,very thick] (21) to node [auto,swap,near start,yshift=0.1cm] {\txt{\textbf{\footnotesize 0}}} (22);
	\end{scope}
	\begin{scope}[xshift=9.4cm,y=1.5cm,yshift=-5cm,x=1cm]
	\node at (.05,0) {\footnotesize matching 0};
	\node at (.05,-.2) {\footnotesize matching 1};
	\node at (.05,-.4) {\footnotesize matching 2};
	\draw[very thick] (1,0)--(1.75,0);
	\draw[thin,dash pattern=on 4pt off 1pt] (1,-.2)--(1.75,-.2);
	\draw[thick,dotted] (1,-.4)--(1.75,-.4);
	\draw (-.9,.2) rectangle (2,-.6);
	\end{scope}
	\end{tikzpicture}}
	\caption{Routing procedure for a 2D nearest-neighbour connectivity. The goal is change the qubit placement depicted in a) to the new qubit placement depicted in d). This routing procedure has three steps, the qubits are numbered 0,...,8 and colored according to which final column they need to be moved. Firstly one permutes qubits within each column, a)$\rightarrow$ b), such that the qubits in each row have a unique color, we explain below how to achieve this. Secondly, one permutes qubits within each row, b)$\rightarrow$ c), so that the qubits are in the correct final column they belong to. Thirdly, one permutes qubits again within each column, c)$\rightarrow$ d), in order to arrive at the final qubit placement. All the permutation within rows and columns are performed in parallel using an odd-even transposition sort \cite{habermann1972parallel} therefore as rows and columns each have $\sqrt{n}$ qubits, this routing is done in worst case with a circuit depth of 3$\sqrt{n}$ steps. The only tricky part of this procedure is step 1, which has been described in \cite{Brierley2017}. One builds a bipartite graph (U,V,E) which has as many nodes $u_i$ and $v_j$ as the square grid has columns, in our case $\sqrt{n}$. An edge is added between node $u_i$ and $v_j$ if there is a qubit initially in column $i$ which ends up in column $j$. This will create a $\sqrt{n}$-regular bipartite graph. Hall's marriage theorem shows that a regular bipartite graph always contains a perfect matching \cite{hall1935}. One finds a perfect matching in this bipartite graph (U,V,E) and labels it 0, then one removes the edges contained in this perfect matching and arrives at a $\sqrt{n}-1$-regular bipartite graph and finds again a perfect matching and labels it 1 and so on until one has $\sqrt{n}$ matchings and no more edges in the graph. Using the found matchings, it is possible to find the desired qubit placement of b) such that qubits in the same row have unique colors. The matching 0 determines which qubits go into row 0 within each column. The matching labeled 0 contains for each column $i$ one edge going from node $u_i$ to some node $v_j$ which means that we move a qubit from within column $i$ to row 0 if it has a final column destination of $j$ (marked by the color). We then continue by assigning elements within each column to row 1 and so on. This then creates the required property that permuting only with each column, we arrive at a qubit placement in b) such that each row contains different colored qubits and the rest of the routing is then trivial.
	}
	\label{fig:2dmapping}
\end{figure*}
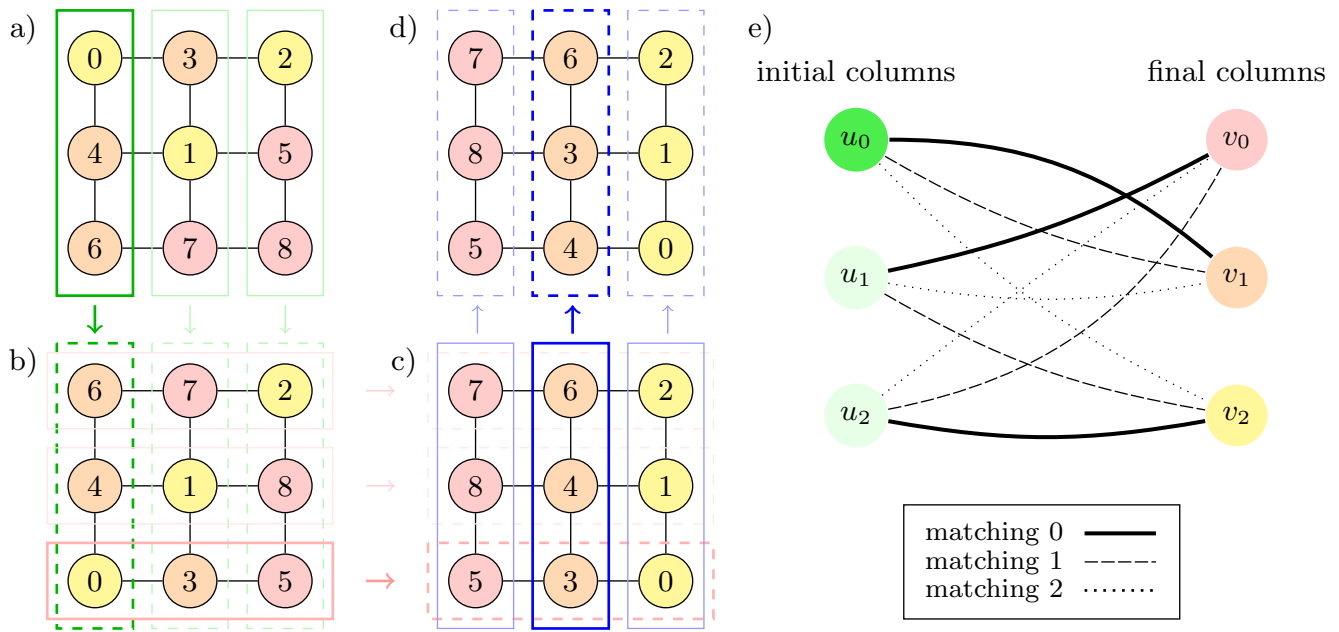

We find the next qubit placement by a greedy search. Our heuristic tries to apply the first commands it encounters by building a linear chain of qubits. For the two dimensional square grid, this linear chain of qubits is then embedded into the two-dimensional grid using a snake pattern.
Our routing schemes are both asymptotically optimal. For a linear chain with $n$ qubits, our scheme uses a standard odd-even transposition sort \cite{habermann1972parallel} which has a worst case circuit depth of $n$ swap operations. For the two-dimensional square grid with $n$ qubits we use the technique of \cite{Brierley2017, Szegedy2017}, which have a worst case depth of 3$\sqrt{n}$ swaps if the grid has equal number of rows and columns (it also works for rectangular grids), see Fig.~\ref{fig:2dmapping} for a graphical explanation of the algorithm.
It is easy to see why (up to constants), the circuit depth overheads of our schemes are optimal as the furthest distance of two qubits on a linear chain is $n$ and on a two-dimensional square grid it is $2\sqrt{n}$.

\begin{figure}[t]
	\resizebox{\linewidth}{!}{\input{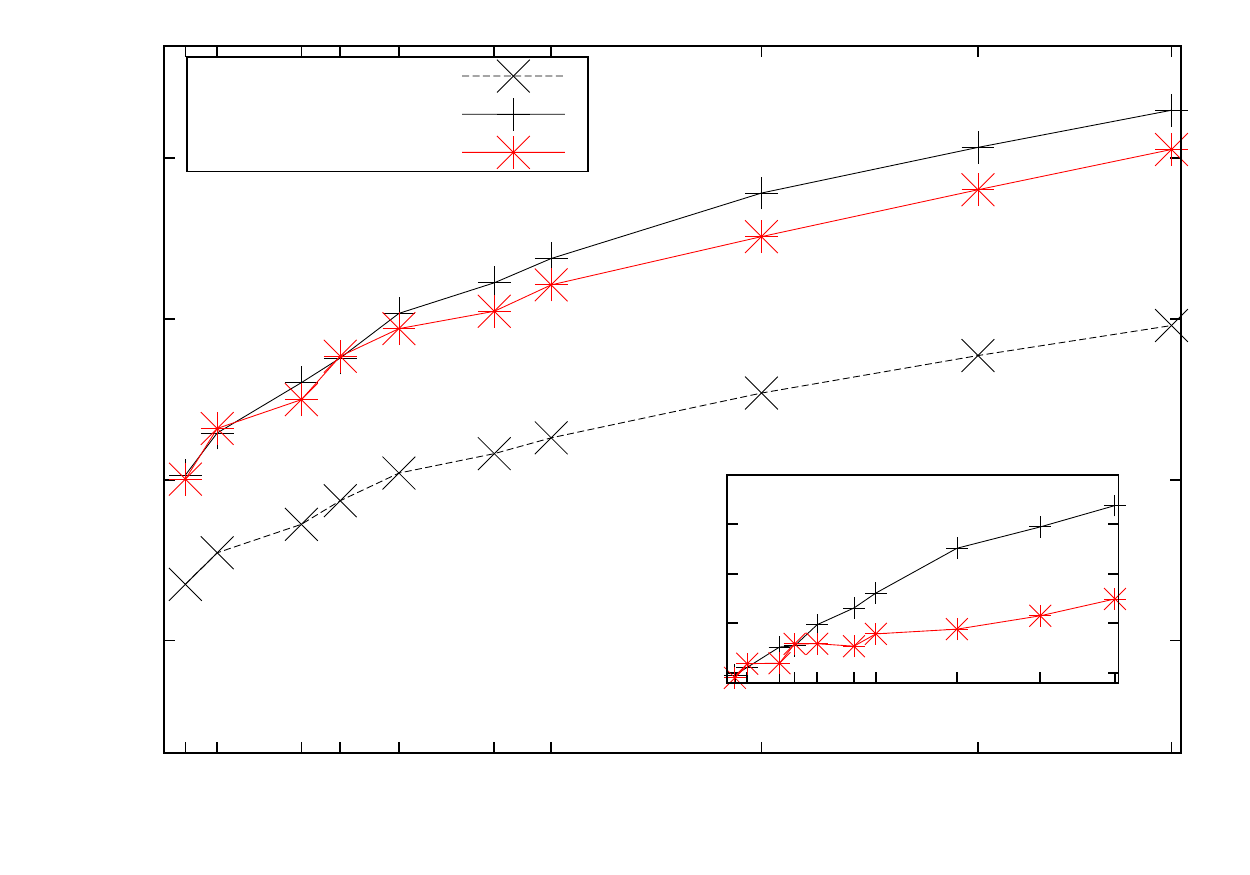}}
	\caption{Comparison between the number of CNOT gates before and after enforcing nearest neighbor connectivity for a 1D and 2D grid.}
	\label{fig:mappingcnot}
\end{figure}

\begin{figure}[t]
	\resizebox{\linewidth}{!}{\input{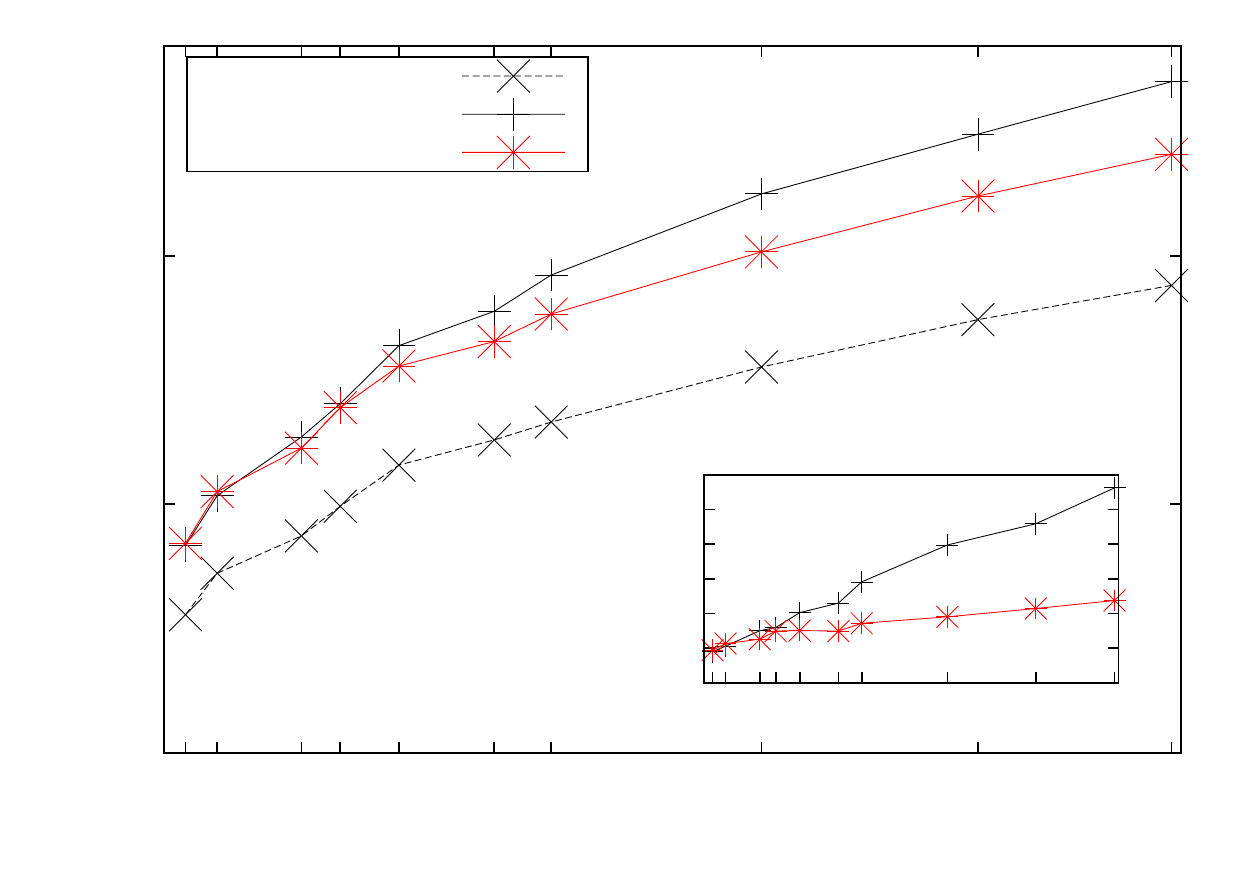}}
	\caption{Comparison between the circuit depth before and after enforcing nearest neighbor connectivity for a 1D and 2D grid.}
	\label{fig:mappingdepth}
\end{figure}

We applied these two mappers to the circuit resulting from the first loop iteration of Shor's algorithm in Box~\ref{box:shor} (for an $n$ bit number, there are $2n$ such iterations which are almost identical). See Fig.~\ref{fig:mappingcnot} and Fig.~\ref{fig:mappingdepth} for a comparison between the circuit depth of an all-to-all connectivity compared to a linear nearest neighbor chain and a two-dimensional square grid. Both of these mappers are available in ProjectQ such that one can improve and test them on more algorithms.

\section{Outlook}
ProjectQ development continues to add and improve features,  in particular more advanced heuristics for the mappers. From a performance point of view, ProjectQ is fast enough for the near term devices but we will continue to move performance heavy tasks to a high-performance C++ implementation. The only current performance bottleneck is estimating resources for large-scale quantum algorithms but this will be solved by introducing a hierarchical resource counter. 
\vfill

\begin{acknowledgments}
We acknowledge support by the Swiss National Science Foundation and the Swiss National Competence Center for Research QSIT. D.S.S. acknowledges very useful discussions about mappers with Mario Szegedy, Harry Buhrman, Stephen Brierley, and Torsten Hoefler.
\end{acknowledgments}

\newpage
\bibliography{references}


\end{document}